\documentclass[twocolumn]{aastex631}
\usepackage{gensymb}
\usepackage{hyperref}
\usepackage{amsmath}
\usepackage{amssymb}
\usepackage{latexsym}
\usepackage{graphicx}
\usepackage{soul}
\usepackage[normalem]{ulem}
\usepackage{color}

\definecolor{notes}{HTML}{5C9384}

\newcommand{\clustername}{SDSS\,J1110$+$6459}
\newcommand{\clustershort}{SDSS\,J1110}
\newcommand{\galaxyname}{SGAS\,J111020.0$+$645950.8}

\newcommand{\lenstool}{\texttt{Lenstool}}

\newcommand{\clustz}{$z$=$0.659$}
\newcommand{\galz}{$z$=$2.481$}

\begin{document}

\title{Precision by Assumption: The Role of Positional Uncertainty in Strong Lens Modeling Posteriors}


\author[0009-0004-9243-3459]{Pedram Abedi}
\affiliation{Department of Astronomy, University of Michigan, 1085 S. University Ave, Ann Arbor, MI 48109, USA}
\correspondingauthor{Pedram Abedi}
\email{pabedi@umich.edu}

\author[0000-0002-7559-0864]{Keren Sharon}
\affiliation{Department of Astronomy, University of Michigan, 1085 S. University Ave, Ann Arbor, MI 48109, USA}

\begin{abstract}
Strong lens models are essential for analyses of magnified background sources and the physical properties of foreground lenses. An estimate of model uncertainties is critical for their accurate interpretations. In this note, we investigate the role of modeling choices on statistical lens model uncertainties. Specifically, we test how the assumed positional uncertainty assigned to image constraints, $\sigma_\mathrm{pos}$, affects derived lensing uncertainties. We find that models with assumed $\sigma_\mathrm{pos}=0\farcs05$, $0\farcs10$, $0\farcs30$, and $0\farcs60$ yield similar best-fit and image-plane RMS values, but substantially different posterior widths that increase with $\sigma_\mathrm{pos}$. Additionally, the uncertainties of some derived quantities are more sensitive than others to this choice. Thus, similar positional agreement does not imply comparable statistical precision. These findings motivate jointly assessing reduced $\chi^2$, image-plane RMS, and changes in derived physical quantities to select a statistically appropriate $\sigma_\mathrm{pos}$. Reporting these diagnostics alongside posterior intervals is essential for interpreting the precision of lens model predictions.

\end{abstract}


\section{Introduction}\label{sec:intro}

Parametric strong gravitational lens models use the positions of multiply imaged features to constrain the foreground mass distribution of the lensing cluster. Within \lenstool\ \citep{Jullo2007}, image-plane optimization minimizes the $\chi^2$ based on positional offsets between observed and model-predicted image locations
\begin{equation}
    \chi^2_{\mathrm{img}} = \sum_{i=1}^{N_{\mathrm{img}}}\frac{\left\|\boldsymbol{x}_{i}^{\mathrm{obs}} -\boldsymbol{x}_{i}^{\mathrm{model}}\right\|^2}{\sigma_{\mathrm{pos}}^2}
    \label{eq:chi2_img}
\end{equation}
where $\boldsymbol{x}_{i}^{\mathrm{obs}}$ and $\boldsymbol{x}_{i}^{\mathrm{model}}$ are the observed and model-predicted positions of each multiply imaged feature, and $\sigma_{\mathrm{pos}}$ is the assumed positional uncertainty assigned to every image. $\sigma_{\mathrm{pos}}$ may accommodate both astrometric errors and deflections from structures absent in the model, and does not need to equal the image-centroid measurement precision \citep[e.g.,][]{Limousin2016}.

The reduced $\chi^2$ statistic assesses the agreement between residuals and assigned errors while accounting for $\nu$ degrees of freedom in the model: $\chi^2_\nu=\chi^2/\nu$. Values near unity indicate statistical consistency; substantially higher values suggest underestimated uncertainties or model inadequacy; substantially lower values indicate overestimated uncertainties or over-fitting. The resulting statistical model uncertainties propagate into derived quantities such as source sizes, luminosities, star-formation rates, time delays, and lens masses. Here, we investigate how $\sigma_{\mathrm{pos}}$ affects the RMS of the best-fit model, and the posterior widths of optimized model parameters, derived magnification, and enclosed-mass measurements. This investigation is restricted to the \lenstool\ modeling framework.

\section{Controlled Model Comparison}

We use the \lenstool\ model of \clustername\ (\clustz; hereafter \clustershort) presented by \citet{abedi2026}. The system comprises four multiply imaged sources, including the giant arc \galaxyname\ at \galz\ -- a highly magnified galaxy with three merging images and $\sim20$ resolved clumps per image. The model includes a cluster-scale dPIE halo, the brightest cluster galaxy, three individually modeled galaxy halos, a cluster member population with parameters scaled to galaxies' luminosities, and external shear. It adopts $\sigma_{\mathrm{pos}}=0\farcs10$ with $\chi^2_\nu=1.16$. See \citet{abedi2026} for a detailed description of the lens model.

We repeat image-plane optimization with $\sigma_{\mathrm{pos}}=0\farcs05$, $0\farcs30$, and $0\farcs60$, retaining the same constraints, mass components, free parameters, and priors. We ensured sufficient MCMC sampling for all runs, establishing convergence within the same region of the parameter space. From \autoref{eq:chi2_img}, we expect $\chi^2\propto\sigma_\mathrm{pos}^{-2}$, with the location of the $\chi^2$ minima unchanged. Larger $\sigma_{\mathrm{pos}}$ values reduce the $\chi^2$ penalty and are therefore expected to permit broader posterior distributions. Assuming approximately Gaussian posterior distributions, negligible influence from priors, and linear propagation to derived quantities, the posterior widths should scale directly with $\sigma_{\mathrm{pos}}$ (see Section 4 of \citealt{hogg2010}). We therefore expect posterior width ratios of 0.5, 3, and 6, respectively, relative to the $0\farcs10$ reference model.

For each run, we evaluate the image- and source-plane RMS of the best-fit model, and the posteriors of optimized parameters and derived quantities. Utilizing 300 random models from the MCMC posterior, we evaluate several lensing outputs: each primary image's areal magnification, $\mu=A_{\mathrm{img}}/A_{\mathrm{src}}$; their summed total magnification; and clump magnifications at observed constraint positions. We also measure the projected mass inside the primary critical curve for \galz\ and circular apertures of $0.5$, $1$, $2$, $3$, $4$, and $5\times$ Einstein radius. Each realization uses its corresponding critical curve and Einstein radius. Cosmology and measurement procedures follow \citet{abedi2026}.

We define the posterior width of each parameter and derived quantity as its 95\% confidence interval. Dividing these widths by the values from the $0\farcs10$ reference model yields the ratios shown in \autoref{fig:uncert}. For optimized model parameters we retain all MCMC samples. 

\section{Results and Implications}

Increasing $\sigma_{\mathrm{pos}}$ yields image-plane RMS values of $0\farcs11$, $0\farcs11$, $0\farcs12$, and $0\farcs15$, and source-plane RMS values $0\farcs049$, $0\farcs048$, $0\farcs048$, and $0\farcs075$. The best-fit values of optimized parameters differ from the $0\farcs10$ reference model by a median of $\sim7\%$. Therefore, both the positional agreement and best-fit parameters remain stable across a twelvefold range in assumed uncertainty.

\begin{figure*}
\centering
    \includegraphics[width=1\linewidth]{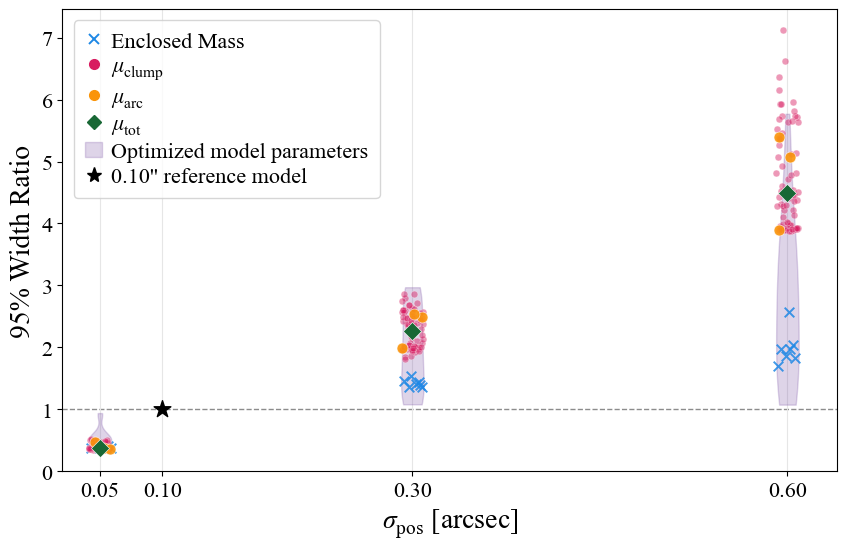}
    \caption{Dependence of statistical uncertainties (encoded as 95\% posterior widths) on $\sigma_{\mathrm{pos}}$ for magnifications, mass, and model parameters. Each width is divided by that of the same quantity in the $0\farcs10$ reference model, shown by the black star. Purple clouds show the distribution of width ratios across all optimized model parameters; wider sections indicate more parameters with similar ratios. Small horizontal offsets were added for clarity, separating overlapping points.} 
    \label{fig:uncert}
\end{figure*}

Despite similar RMS and best-fit values, the posterior widths change substantially (\autoref{fig:uncert}). Halving $\sigma_\mathrm{pos}$ narrows both magnification and enclosed-mass widths by $\gtrsim2\times$, while model-parameter widths narrow by $\sim2\times$. Increasing $\sigma_\mathrm{pos}$ sixfold broadens magnification widths by $\sim4.5\times$, compared with $\sim2\times$ for enclosed-mass and $\sim2.5\times$ for model parameters. These different responses demonstrate that changing $\sigma_\mathrm{pos}$ does not uniformly rescale uncertainties across model parameters or derived quantities.

The width ratios depart from the linear expectation of 0.5, 3, and 6. These deviations may arise from prior boundaries, non-Gaussian posterior structure, or nonlinear propagation from model parameters to derived quantities. Prior boundaries can limit how much parameter intervals broaden as $\sigma_{\mathrm{pos}}$ increases, causing the width ratios to plateau at larger $\sigma_{\mathrm{pos}}$. Additionally, although the posterior structure of a free parameter appears Gaussian, correlations with other parameters and their priors can drive nonlinear scaling. Magnifications and enclosed-masses are derived from these fitted parameters, inheriting both the prior bounds and underlying parameter degeneracies.

In strong-lens modeling, degeneracies allow halo parameters to compensate for one another while preserving the observed image positions, affecting magnification and enclosed-mass differently. Magnification is sensitive to the spatial distribution of mass, particularly near critical curves. Enclosed-mass sums the projected mass within an aperture: redistributing mass inside that boundary can alter local magnifications without changing the enclosed-mass. This provides a possible explanation for the weaker broadening of enclosed-mass intervals compared with magnification intervals.

For \lenstool, and likely other parametric algorithms, these results emphasize that $\sigma_\mathrm{pos}$ is an important factor in interpreting a model's statistical precision and posterior distributions, especially when combining or comparing results from different models. Modelers should test plausible values of $\sigma_\mathrm{pos}$ and examine their effects on derived physical quantities. The $\chi^2_\nu$ provides a useful consistency check; values substantially above or below unity motivate investigating the mass model or assigned uncertainties. An uncertainty comparable to the image-plane RMS offers a practical starting point. For example, in \clustershort, $\sigma_\mathrm{pos}=0\farcs10$ is comparable to the $0\farcs11$ RMS and yields $\chi^2_\nu=1.16$, supporting its adoption for this configuration. Reporting $\sigma_\mathrm{pos}$, $\chi^2_\nu$, and RMS alongside posterior intervals makes the inferred precision interpretable, helps identify when further modeling work is needed, and enables statistically-meaningful comparisons between models.

\vspace{5mm}
\begin{acknowledgements} 

The authors wish to thank the anonymous referee of \citet{abedi2026} for useful comments prompting this investigation.

Support for JWST programs GO-03843, GO-04125 was provided by NASA through a grant from STScI, which is operated by AURA under NASA contract NAS5-03127.

\end{acknowledgements}

\bibliography{masterbib}{}
\bibliographystyle{aasjournal}

\end{document}